# Direct Measurement of the Critical Cooling Rate for the Vitrification of Water

Nathan J. Mowry[†], Constantin R. Krüger[†], Marcel Drabbels, and Ulrich J. Lorenz[*]

**Affiliation:** Ecole Polytechnique Fédérale de Lausanne (EPFL), Laboratory of Molecular Nanodynamics, CH-1015 Lausanne, Switzerland

[†] These authors contributed equally

[*] To whom correspondence should be addressed. Email: ulrich.lorenz@epfl.ch

**Abstract**

The vitrification of aqueous solutions through rapid cooling is a remarkable achievement that launched the field of cryo-electron microscopy (cryo-EM) and has enabled the cryopreservation of biological specimens. For judging the feasibility of a vitrification experiment, the critical cooling rate of pure water is a frequently cited reference quantity. However, an accurate determination has remained elusive, with estimates varying by several orders of magnitude. Here, we employ *in situ* and time-resolved electron microscopy to obtain a precise measurement. We use shaped microsecond laser pulses to briefly melt an amorphous ice sample before flash freezing it with a variable, well-defined cooling rate. This allows us to directly measure the critical cooling rate of pure water, which we determine to be $6.4 \cdot 10^6$ K/s. Our experimental approach also expands the toolkit of microsecond time-resolved cryo-EM, an emerging technique, in which a cryo sample is similarly flash melted and revitrified with a laser pulse.

For a long time, it was believed that the vitrification of water would require excessively high cooling rates and was therefore impossible.[1] Indeed, upon supercooling, water enters the so-called no man's land (160–232 K),[2] where crystallization occurs within tens of microseconds.[3] However, if no man's land can be traversed rapidly enough to outpace crystallization, hyperquenched glassy water (HGW) is obtained, a type of amorphous ice with a glass transition temperature of 136 K.[4] Sufficiently high cooling rates to achieve vitrification were first realized by injecting a thin water jet into a cryogenic liquid[5] and by plunging small water droplets on a thin specimen support into a cryogen.[6] Other methods that have since been demonstrated include impinging small droplets on a cold surface[7] and directing a jet of cryogenic liquid at an aqueous sample.[8]

In order to assess whether a given method is suitable to achieve vitrification, its cooling speed is frequently compared to the critical cooling rate of pure water, which for most samples represents an upper limit for the required cooling rate. This is because the critical cooling rate of aqueous solutions decreases exponentially with increasing solute concentration.[4] However, an accurate measurement of the critical cooling rate of pure water has remained elusive, with estimates ranging from $10^5$ K/s to $10^{12}$ K/s (see Ref. 9 for an overview). A rate of $3 \cdot 10^5$ K/s has been estimated by extrapolating from aqueous solutions with molar solute concentrations.[4] However, it is not clear whether such an extrapolation to zero concentration is accurate. Experiments on microdroplets that are vitrified by impinging on a cold surface yield a critical cooling rate of $10^6$–$10^7$ K/s.[7,10] However, in such experiments, a distribution of cooling rates is likely to occur across the droplet volume, with some droplets containing a small fraction of crystalline ice even at high cooling rates.[10]

An accurate determination of the critical cooling rate of pure water would require an experiment in which sufficiently high cooling rates can be realized and systematically varied and in which the spread of cooling rates across the sample volume is limited. Here, we use *in situ* and time-resolved electron microscopy[11,12] to implement such an experiment, building on methodology that we have previously introduced.[3,13] We use shaped microsecond laser pulses[14] to rapidly melt a thin amorphous ice sample *in situ* before flash freezing it with a well-controlled cooling rate. We then record an electron diffraction pattern to determine whether crystallization has occurred.

Figure 1a,b illustrates our experimental approach (Supporting Information 1).[3,11,13] A few-layer graphene sheet serves as the sample support, which is suspended over a holey gold film (2 µm holes) on a 600 mesh gold grid that is held at 100 K (Fig. 1a). We deposit a 206 nm thick layer of amorphous solid water (ASW) by leaking water vapor into the sample region of our electron microscope through a gas dosing valve and then use a 30 µs laser pulse (532 nm) to locally melt the ASW film in the center of a grid square. At the end of the laser pulse, the sample reaches a temperature of 294±6 K, as determined from time-resolved diffraction patterns, which we record with 10 µs electron pulses.[3,11,12] Once the laser is switched off, the sample cools rapidly as the heat is dissipated towards the surroundings, which have remained at cryogenic temperature. If we instead reduce the laser power more gradually, we are able to flash freeze the sample with a variable cooling rate. We then record a diffraction pattern in order to probe for the formation of crystals (Fig. 1b). Finally, we evaporate the sample through laser heating and deposit a fresh layer of ASW to repeat the experiment.

Figure 1c displays the laser pulse shapes used as measured with a photodiode, with simulations in Fig. 1d illustrating the corresponding temperature evolution of the sample. After 30 µs, we immediately drop the laser power by 20 % and then linearly ramp it down to zero. This particular pulse shape allows us to achieve faster initial cooling and reduce sample evaporation. The simulations reveal that at any given point in time, the spread of cooling rates across the sample volume probed by the electron beam is small, with a standard deviation of less than 1.6 %. Note that as discussed in more detail below, we ultimately do not rely on simulations to obtain the critical cooling rate of the sample, but instead determine it experimentally with a time-resolved electron diffraction experiment. In our simulations, we have adjusted the characteristic cooling time of the sample to match the experiment (Supporting Information 2).

Typical diffraction patterns of samples flash frozen at different cooling rates are shown in Fig. 2a, with the corresponding azimuthally averaged patterns in Fig. 2b. For each panel, the time interval is indicated over which the laser power is ramped to zero. At the highest cooling rates, crystallization is avoided entirely, and all the diffraction patterns obtained are indistinguishable from that of HGW. As the cooling rate is decreased, intense diffraction features appear in some experiments, which indicate the formation of hexagonal ice.

We estimate the crystalline fraction of the sample by decomposing the diffraction patterns into contributions of HGW and hexagonal ice (Fig. 3a, purple and green curves, respectively). The diffraction pattern of HGW was recorded from samples flash frozen with the maximum cooling rate we can achieve in our experiment, *i.e.* the laser is switched off immediately after 30 µs, while the diffraction pattern of hexagonal ice was obtained from samples that were crystallized by laser heating them to a temperature of about 265 K (Supporting Information 3). Figure 3b confirms that the experimental diffraction patterns (thin lines, averages of 9–14 experiments, Supporting Information 1) can be reasonably well reproduced by weighted sums of these two components (thick lines), where we have restricted the sum of the weights to one. Figure 3c displays the component weights as a function of the time interval over which the laser power is ramped to zero. As the sample is cooled more slowly, the weight of the crystalline component suddenly rises steeply as crystallization sets in. At the same time, the fraction of experiments increases, in which crystallization occurs (Fig. 3d). We find that a cooling ramp duration of 24 µs (marked with an arrow) yields the lowest cooling rate for which no crystallization occurs and which we therefore identify as the critical cooling rate.

We obtain an accurate measurement of critical cooling rate from a time-resolved electron diffraction experiment (Supporting Information 1), which allows us to determine the speed at which the sample cools during a 24 µs cooling ramp. Figure 4a displays the temporal evolution of the first diffraction maximum as captured with 2 µs electron pulses (Fig. 4a). For comparison, Fig. 4b shows the temporal evolution at the maximum cooling rate achievable in our experiment. As the sample cools, the first diffraction maximum shifts to lower momentum transfer, until its position converges to that of HGW (horizontal dashed line). As we have previously shown, this occurs at a temperature of just below 200 K.[3] We fit the position of the first diffraction maximum (black dots) with a sigmoid (solid line). From this fit, we can then extract the temperature evolution of the sample (Fig. 4c), since we have previously established the temperature dependence of the peak position throughout no man's land[3] (Supporting Information 4).

We determine the critical cooling rate from the temperature evolution of the sample as follows. Figure 4c shows that in the range of 200–240 K, the cooling rate of the sample is nearly constant, with a standard deviation of 7 %. Simulations of the crystallization kinetics[13] during flash freezing show that in the range of cooling rates studied here, crystallization predominantly occurs in this temperature interval (Supporting Information 5). The temperature evolution of the sample outside of this range can therefore be neglected for the determination of the critical cooling rate. This is because at higher temperatures, the nucleation rate is too low for any significant amount of crystallization to occur on the timescale of our experiment, while at lower temperatures, the growth rate decreases exponentially, so that crystal growth ceases.[2] From a linear fit of the temperature evolution between 200 K and 240 K (black dashed line), we obtain an average cooling rate of $1.1\pm0.1\cdot10^7$ K/s for an experiment in which the laser is switched off immediately after 30 µs. For the critical cooling rate, we obtain a value of $6.4\pm0.5\cdot10^6$ K/s.

In conclusion, we have obtained a direct measurement of the critical cooling rate of pure water. In our 206 nm thick samples, surface nucleation is likely to play a role,[15,16] although it is not clear whether the critical cooling rate would be higher or lower in a bulk sample. It was found that surface nucleation dominates the crystallization process in very thin amorphous ice samples. Crystallization times increase with sample thickness and plateau between 26–55 nm, but grow shorter again as the sample thickness is increased further.[17]

The critical cooling rate determined here is more than one order of magnitude lower than the critical heating rate for outrunning crystallization during flash melting of amorphous ice samples of $10^8$ K/s, which we have previously determined with time-resolved electron diffraction experiment in a near-identical sample geometry.[14] This is qualitatively consistent with previous studies, which found that for solutions of cryoprotectants, the critical heating rate is usually one to two orders of magnitude higher than the critical cooling rate.[18,19] This is a consequence of the different temperature dependence of the nucleation and growth rates, as has been previously pointed out.[20] During flash melting, the sample first traverses a temperature range, in which fast nucleation occurs. While nucleation slows dramatically as the sample reaches higher temperatures, the growth rate increases exponentially, which causes the sample to crystallize. During flash freezing, the sequence of events is reversed. Large concentrations of nuclei are only formed once the sample has cooled to low temperatures, where the growth of these nuclei is slow.

Our experiments also broaden the toolkit of microsecond time-resolved cryo-EM experiments, in which a cryo sample is flash melted with a laser beam to briefly allow dynamics of the embedded proteins to occur, before the laser is switched off, and the sample is revitrified.[21,22] By shaping the laser pulses as we have done here, it becomes possible to systematically investigate the effect of the cooling rate on cryo sample preparation. For example, the vitrification speed has been found to affect the sample drift observed during imaging with the electron beam,[23,24] with lower cooling rates reducing the beam-induced motion.[23] Systematically varying the cooling rate will also make it possible to experimentally investigate how the vitrification process affects protein conformational distributions and whether it may induce structural transitions in some particles.[25]

**Supplementary Material:**

Experimental methods, simulation of the temperature evolution of the sample, decomposition of the diffraction patterns into contributions of HGW and hexagonal ice, determination of the temperature evolution of the sample at the critical cooling rate, and simulation of the crystallization kinetics

**Acknowledgments:**

This work was supported by the Swiss National Science Foundation Grant 200020_207842.

**Author contributions:**

Conceptualization: UJL

Methodology: NJM, CRK, UJL

Investigation: NJM, CRK, UJL

Visualization: NJM, CRK, UJL

Funding acquisition: UJL

Project administration: UJL

Supervision: UJL

Writing – original draft: NJM, CRK, UJL

Writing – review & editing: NJM, CRK, MD, UJL

**Competing interests:**

The authors declare that they have no competing interests.

**Data and materials availability:**

The data and materials can be accessed on Zenodo: DOI to be added

**References**

(1) Dubochet, J. Cryo-EM—the First Thirty Years. *J. Microsc.* **2012**, *245* (3), 221–224.

(2) Gallo, P.; Amann-Winkel, K.; Angell, C. A.; Anisimov, M. A.; Caupin, F.; Chakravarty, C.; Lascaris, E.; Loerting, T.; Panagiotopoulos, A. Z.; Russo, J.; Sellberg, J. A.; Stanley, H. E.; Tanaka, H.; Vega, C.; Xu, L.; Pettersson, L. G. M. Water: A Tale of Two Liquids. *Chem. Rev.* **2016**, *116* (13), 7463–7500.

(3) Krüger, C. R.; Mowry, N. J.; Bongiovanni, G.; Drabbels, M.; Lorenz, U. J. Electron Diffraction of Deeply Supercooled Water in No Man's Land. *Nat. Commun.* **2023**, *14* (1), 2812.

(4) Warkentin, M.; Sethna, J. P.; Thorne, R. E. Critical Droplet Theory Explains the Glass Formability of Aqueous Solutions. *Phys. Rev. Lett.* **2013**, *110* (1), 015703.

(5) Brüggeller, P.; Mayer, E. Complete Vitrification in Pure Liquid Water and Dilute Aqueous Solutions. *Nature* **1980**, *288* (5791), 569–571.

(6) Dubochet, J.; McDowall, A. w. Vitrification of Pure Water for Electron Microscopy. *J. Microsc.* **1981**, *124*, 3–4.

(7) Kohl, I.; Bachmann, L.; Hallbrucker, A.; Mayer, E.; Loerting, T. Liquid-like Relaxation in Hyperquenched Water at ≤140 K. *Phys. Chem. Chem. Phys.* **2005**, *7* (17), 3210.

(8) Ravelli, R. B. G. Cryo-EM Structures from Sub-Nl Volumes Using Pin-Printing and Jet Vitrification. *Nat Commun* **2020**, *11*, 2563.

(9) Engstrom, T.; Clinger, J. A.; Spoth, K. A.; Clarke, O. B.; Closs, D. S.; Jayne, R.; Apker, B. A.; Thorne, R. E. High-Resolution Single-Particle Cryo-EM of Samples Vitrified in Boiling Nitrogen. *IUCrJ* **2021**, *8* (6), 867–877.

(10) Bachler, J.; Giebelmann, J.; Loerting, T. Experimental Evidence for Glass Polymorphism in Vitrified Water Droplets. *Proc. Natl. Acad. Sci.* **2021**, *118* (30), e2108194118.

(11) Olshin, P. K.; Bongiovanni, G.; Drabbels, M.; Lorenz, U. J. Atomic-Resolution Imaging of Fast Nanoscale Dynamics with Bright Microsecond Electron Pulses. *Nano Lett.* **2021**, *21* (1), 612–618.

(12) Bongiovanni, G.; Olshin, P. K.; Drabbels, M.; Lorenz, U. J. Intense Microsecond Electron Pulses from a Schottky Emitter. *Appl. Phys. Lett.* **2020**, *116* (23), 234103.

(13) Mowry, N. J.; Krüger, C. R.; Bongiovanni, G.; Drabbels, M.; Lorenz, U. J. Flash Melting Amorphous Ice. *J. Chem. Phys.* **2024**, *160* (18), 184502.

(14) Krüger, C. R.; Mowry, N. J.; Drabbels, M.; Lorenz, U. J. Shaped Laser Pulses for Microsecond Time-Resolved Cryo-EM: Outrunning Crystallization during Flash Melting. *J Phys Chem Lett* **2024**, *15*, 4244−4248.

(15) Backus, E. H. G.; Grecea, M. L.; Kleyn, A. W.; Bonn, M. Surface Crystallization of Amorphous Solid Water. *Phys. Rev. Lett.* **2004**, *92* (23), 236101.

(16) Yuan, C.; Smith, R. S.; Kay, B. D. Surface and Bulk Crystallization of Amorphous Solid Water Films: Confirmation of "Top-down" Crystallization. *Surf. Sci.* **2016**, *652*, 350–354.

(17) Tonauer, C. M.; Fidler, L.-R.; Giebelmann, J.; Yamashita, K.; Loerting, T. Nucleation and Growth of Crystalline Ices from Amorphous Ices. *J. Chem. Phys.* **2023**, *158* (14), 141001.

(18) Boutron, P.; Mehl, P. Theoretical Prediction of Devitrification Tendency: Determination of Critical Warming Rates without Using Finite Expansions'. *Cryobiology* **1990**, *27*, 359–377.

(19) Han, Z.; Bischof, J. C. Critical Cooling and Warming Rates as a Function of CPA Concentration. *CryoLetters* **2020**, *41* (4), 185–193.

(20) Hopkins, J. B.; Badeau, R.; Warkentin, M.; Thorne, R. E. Effect of Common Cryoprotectants on Critical Warming Rates and Ice Formation in Aqueous Solutions. *Cryobiology* **2012**, *65* (3), 169–178.

(21) Harder, O. F.; Barrass, S. V.; Drabbels, M.; Lorenz, U. J. Fast Viral Dynamics Revealed by Microsecond Time-Resolved Cryo-EM. *Nat. Commun.* **2023**, *14* (1), 5649.

(22) Lorenz, U. J. Microsecond Time-Resolved Cryo-Electron Microscopy. *Curr. Opin. Struct. Biol.* **2024**, *87*, 102840.

(23) Wu, C.; Shi, H.; Zhu, D.; Fan, K.; Zhang, X. Low-Cooling-Rate Freezing in Biomolecular Cryo-Electron Microscopy for Recovery of Initial Frames. *QRB Discov.* **2021**, *2*, e11.

(24) Harder, O. F.; Voss, J. M.; Olshin, P. K.; Drabbels, M.; Lorenz, U. J. Microsecond Melting and Revitrification of Cryo Samples: Protein Structure and Beam-Induced Motion. *Acta Crystallogr. Sect. Struct. Biol.* **2022**, *78* (7), 883–889.

(25) Bock, L. V.; Grubmüller, H. Effects of Cryo-EM Cooling on Structural Ensembles. *Nat. Commun.* **2022**, *13* (1), 1709.

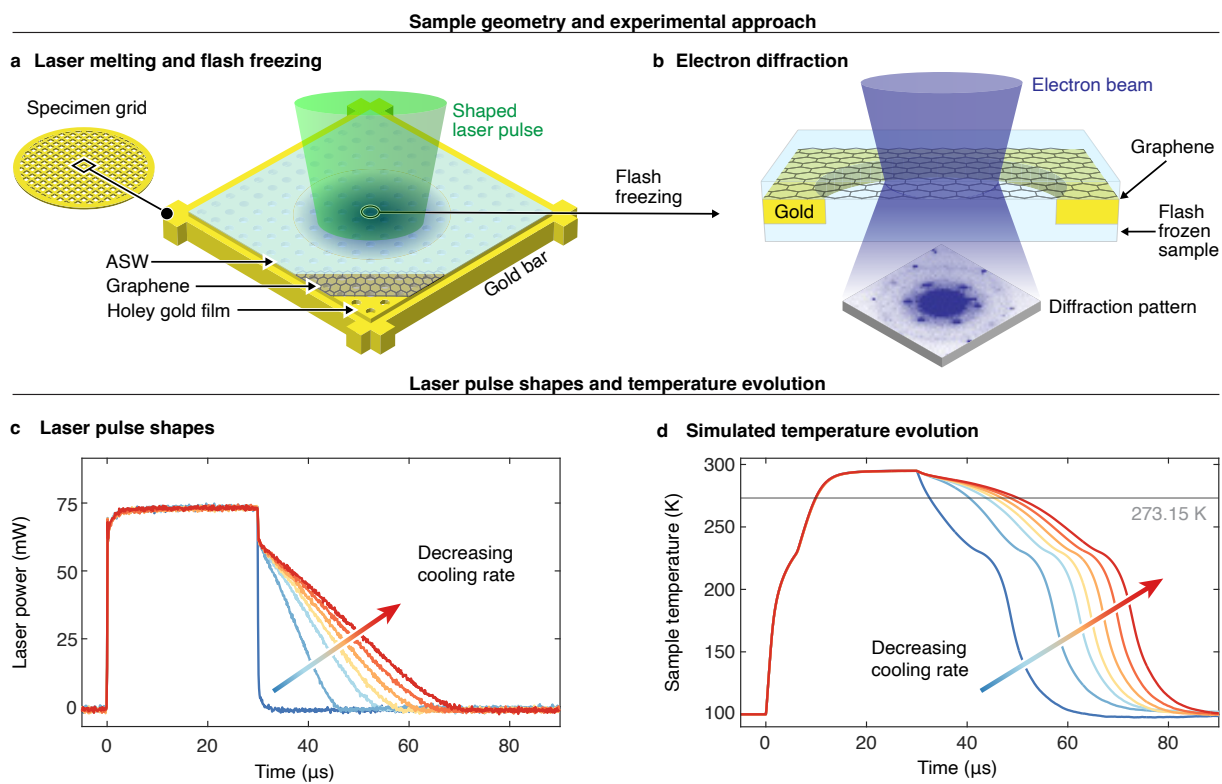

**Figure 1 | Illustration of the experimental approach and simulation of the temperature evolution of the sample under irradiation with shaped laser pulses. a,b** Sample geometry and experimental approach. A 206 nm thick layer of ASW is deposited onto multilayer graphene on a holey gold specimen grid that is cooled to 100 K (**a**). A shaped microsecond laser pulse is then used to briefly melt the sample and flash freeze it with a variable cooling rate, after which an electron diffraction pattern is recorded (**b**). **c** Laser pulse shapes as captured with a photodiode. After 30 μs, the laser power is reduced by 20 %, after which it is linearly decreased to zero. **d** Simulated temperature evolution of the sample under irradiation with the shaped laser pulses in **c**.

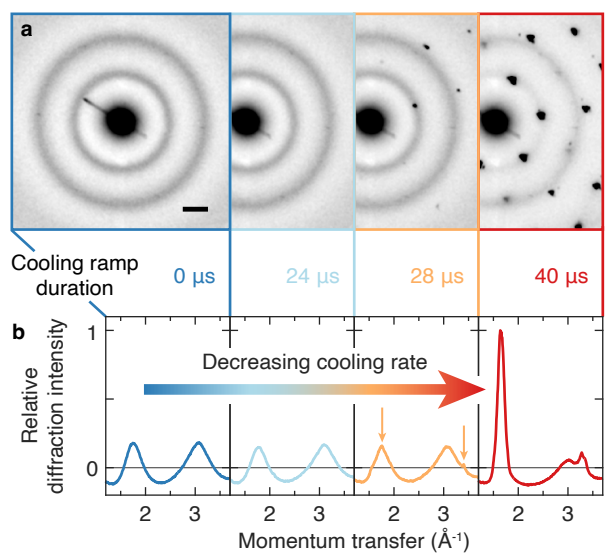

**Figure 2 | Typical diffraction patterns of samples flash frozen with different cooling rates. a** As the cooling rate is decreased, *i.e.* the time interval is increased over which the laser power is ramped to zero (indicated), diffraction peaks appear in some experiments, indicating that crystallization has occurred. At the highest cooling rates, the diffraction patterns are indistinguishable from that of HGW. Note that the weak diffraction spots that remain visible at these rates arise from the graphene support of the sample. The diffraction patterns are shown with a Gaussian filter applied. Scale bar, 1 Å$^{-1}$. **b** Azimuthal averages of the diffraction patterns in **a**. Arrows highlight crystalline diffraction features.

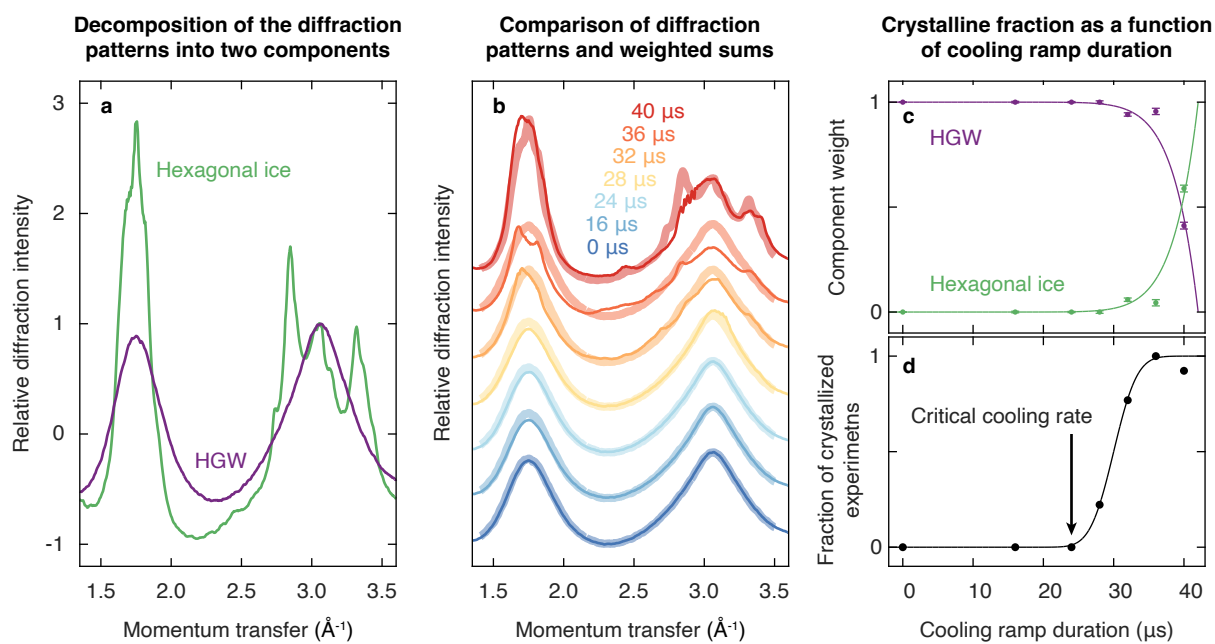

**Figure 3 | Crystalline fraction of the sample as a function of the cooling rate. a** The diffraction patterns of samples flash frozen with different cooling rates (Fig. 2) contain contributions from HGW (purple) and hexagonal ice (green). **b** Weighted sums of the components in **a** (thick lines) reproduce experimental diffraction patterns well (thin lines). The time interval is indicated over which the laser power is ramped to zero. **c** Weight of the HGW and hexagonal ice components as a function of the cooling ramp duration. Error bars represent standard errors of the fit. The solid lines provide a guide to the eye and are derived from a fit with a power function. **d** Fraction of experiments in which crystallization occurs as a function of the cooling ramp duration. The solid line serves as a guide to the eye and is derived from a fit with an error function.

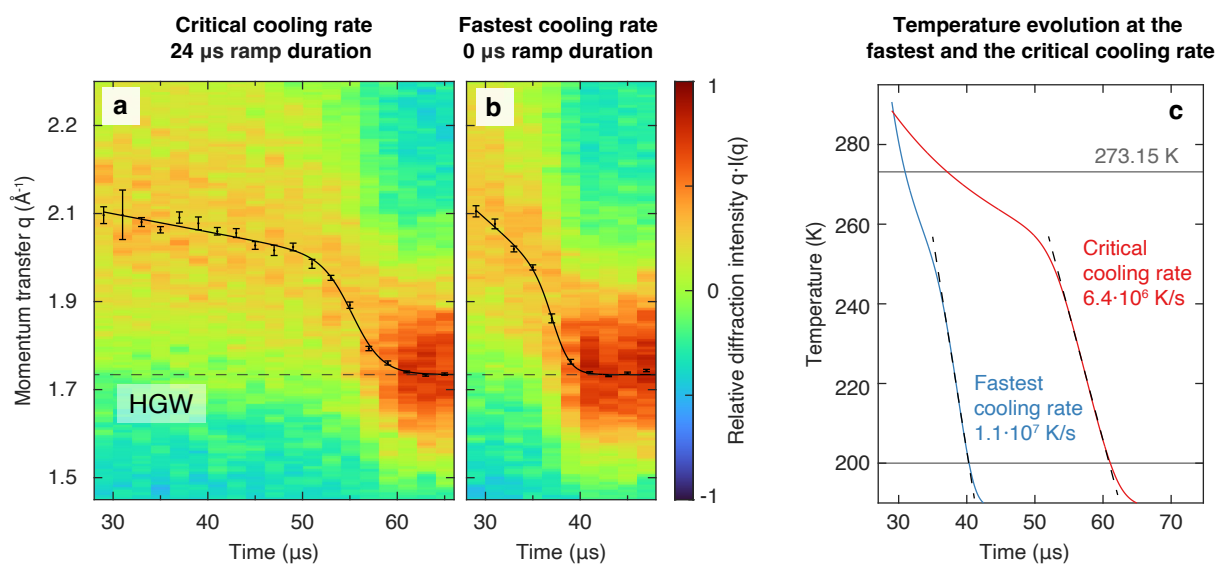

**Figure 4 | Determination of the critical cooling rate from the structural evolution of the sample as observed with time-resolved electron diffraction. a** Structural evolution of the sample at the critical cooling rate. After 30 μs, the laser power is ramped to zero over a time interval of 24 μs, and time-resolved diffraction patterns are captured with 2 μs electron pulses. The position of the first diffraction maximum (black dots) is determined from polynomial fits, with error bars derived from the errors of the fit parameters.[3] The solid line represents a fit of the peak position with a sigmoid, and the horizontal dashed line indicates the peak position for HGW. **b** For comparison, the structural evolution of the sample is shown at the highest cooling rate achievable in our experiment, which we obtain when the laser is immediately switched off after 30 μs. **c** Temperature evolution of the samples in **a** and **b** as determined from the position of the first diffraction maximum. Linear fits in the range of 200–240 K (dashed lines) yield cooling rates of $6.4\pm0.5\cdot10^6$ K/s and $1.1\pm0.1\cdot10^7$ K/s, respectively.

# Supplementary Information

# Direct Measurement of the Critical Cooling Rate for the Vitrification of Water

Nathan J. Mowry[†], Constantin R. Krüger[†], Marcel Drabbels, and Ulrich J. Lorenz[*]

**Affiliation:** Ecole Polytechnique Fédérale de Lausanne (EPFL), Laboratory of Molecular Nanodynamics, CH-1015 Lausanne, Switzerland

**This PDF file includes:**

1  Experimental methods

2  Simulation of the temperature evolution of the sample

3  Decomposition of the diffraction patterns into contributions of HGW and hexagonal ice

4  Determination of the temperature evolution of the sample at the critical cooling rate

5  Simulation of the crystallization kinetics

† These authors contributed equally

* To whom correspondence should be addressed. Email: ulrich.lorenz@epfl.ch

# 1 Experimental methods

The experimental setup, sample geometry, and data analysis procedure have been detailed previously.[1–5] Experiments are performed with a JEOL 2010F transmission electron microscope that we have modified for *in situ* and time-resolved experiments.[1,2] A Quantifoil (Au) R2/1 (N1-A15nAu60-50) grid (50 nm thick holey gold film with 2 µm diameter holes, 1 µm apart on 600 mesh gold) serves as the specimen support, onto which a sheet of 6-8 layer graphene is transferred.[3–5] A cryo specimen holder is used to cool the sample to 100 K, and a 206 nm thick layer of amorphous solid water (ASW) is deposited *in situ* by leaking water vapor into the vacuum of the electron microscope through a gas dosing valve.[3,4] We then melt the sample in the center of a grid square with a shaped 30 µs laser pulse (532 nm, 73.2 mW, 38 µm FWHM spot size in the sample plane), which we obtain by modulating the output of a continuous wave laser with an acousto-optic modulator. In order to determine the sample temperature at the end of the 30 µs laser pulse, we capture a diffraction pattern from the central hole of the grid square with a 10 µs electron pulse,[2,3] where we converge the electron beam to a disc of about 1.5 µm diameter. After 30 µs, we immediately reduce the laser power by 20 % and then gradually ramp it to zero (pulse shapes in Fig. 1c recorded with a photodiode, averages of 10), with the slope of the ramp determining the cooling rate of the sample. In order to determine whether crystallization has occurred, a static diffraction pattern of the sample is captured (20 electron pulses of 10 µs duration at 10 Hz repetition rate). Table S1 provides an overview of how many experiments were performed for each cooling rate, and how many times crystallization was observed. Finally, the sample is evaporated through laser irradiation, and a fresh layer of ASW is deposited to repeat the experiment. The time-resolved diffraction experiments in Fig 4 a,b are performed as previously described[4] with 2 µs electron pulses.[2,3]

| Cooling ramp duration (µs) | 0 | 16 | 24 | 28 | 32 | 36 | 40 |
|---|---|---|---|---|---|---|---|
| Number of experiments | 14 | 14 | 13 | 9 | 13 | 13 | 13 |
| Number of times crystallization was observed | 0 | 0 | 0 | 2 | 10 | 13 | 12 |

**Table S1 | Number of experiments performed for each cooling rate together with the number of experiments in which crystallization was observed.**

**2 Simulation of the temperature evolution of the sample**

We simulate the temperature evolution of the sample under laser irradiation with different pulse shapes (Fig. 1d) with COMSOL Multiphysics 6.1.[3–5] We use the experimentally determined laser pulse shapes (Fig. 1c) and adjust the laser power such that during the 30 μs laser pulse, the sample plateaus at 294 K, the temperature we measure in our experiment. In Fig. 1d, we report the average temperature of the sample within the cylindrical volume of 1.5 μm diameter that is probed by the electron beam.

As shown in Fig. 4c, we determine that at the critical cooling rate, the sample reaches a temperature of 200 K at about 61 μs. When we use the same simulation parameters as previously,[3–5] we find that the sample cools somewhat too rapidly, indicating that the idealized geometry in the simulation slightly deviates from the experiment. For example, it is conceivable that the holey gold film does not have a perfect thermal contact with the gold bars of the specimen grid, as we assume in our simulation, which slows heat dissipation in the experiment. We account for this discrepancy in a simple manner by reducing the thickness of the holey gold foil to 21.1 nm, which slows heat dissipation. With this adjustment, the sample cools to a temperature of 200 K at 61 μs, as observed in the experiment. For comparison, if we simulate the flash freezing process with the fastest cooling time achievable in our experiment, *i.e.* the laser is immediately switched off after 30 μs, the sample reaches 200 K at 47.5 μs. This is about 60 % slower than observed experimentally (41 μs, Fig. 4c), showing that our adjustment of the heat transfer properties of the sample geometry does not work equally well for all cooling rates.

**3 Decomposition of the diffraction patterns into contributions of HGW and hexagonal ice**

In order to estimate the crystalline fraction of the sample after flash freezing with different cooling rates, we decompose the diffraction patterns into contributions of hexagonal ice and HGW (Fig. 3a). The diffraction pattern of HGW was obtained from samples cooled at the maximum cooling rate achievable in our experiment (1.1±0.1·10$^7$ K/s) and represents an average of 4 experiments. To correct for variations in sample thickness and in the intensity of the electron beam, the intensity of the diffraction pattern was slightly scaled to maximize agreement with the diffraction pattern of the sample obtained at the highest cooling rate in Fig. 3b. The diffraction pattern of hexagonal ice was obtained by laser heating freshly deposited ASW samples to a temperature of about 230 K with a 30 μs laser pulse in order to induce the formation of a large number of crystallites. The samples were then heated to a temperature of about 265 K through irradiation with a second, more intense laser pulse (30 µs duration) to transform any stacking disordered ice crystallites into hexagonal ice. Irradiation of the sample with further laser pulses does not increase the intensity of the crystalline diffraction features, showing that the samples have fully crystallized. The diffraction pattern of hexagonal ice shown in Fig. 3a represents an average of 4 experiments. To correct for variations in sample thickness and in the intensity of the electron beam, the intensity of the diffraction pattern was slightly scaled, so that the diffraction intensity below 1 Å$^{-1}$ matches that of the diffraction pattern of HGW in Fig. 3a.

**4 Determination of the temperature evolution of the sample at the critical cooling rate**

We use the following procedure to obtain the temperature evolution of the sample from the time-resolved diffraction experiments in Fig. 4a, b. We fit the position of the first diffraction maximum $q$ with a sigmoid

$$q\,(t) = \frac{b + c * t}{1 + e^{a*(t-t_0)}} + q_{HGW}$$

where $a$, $b$, $c$, and $t_0$ are fit parameters, $q_{HGW}$ is the peak position of HGW, and $t$ is the time. We then extract the temperature evolution of the sample (Fig. 4c) by comparing the fit with the temperature dependence of the peak position that we have previously established,[3] which we fit with another sigmoid as described in equation 3 of Ref. 6.

**5 Simulation of the crystallization kinetics**

We simulate the crystallization kinetics of the sample in order to determine the temperature interval, in which crystallization predominantly occurs and which is therefore relevant for the determination of the critical cooling rate. Figure S1 displays a simulation of the crystalline fraction of the sample as a function of temperature for different cooling rates. Details of the simulations have been described previously.[3,4] Since experimental nucleation rates of water have uncertainties of several orders of magnitude in no man's land,[7–13] we use a simple model for the nucleation rate, which we describe as before with a super-Gaussian with a maximum rate of 3.4·10$^{26}$ m$^{-3}$s$^{-1}$ in no man's land.[4] With this choice, we have previously been able to model the crystallization kinetics of amorphous ice samples during flash melting.[4] The simulations in Fig. S1 show that crystallization only sets in below 225 K and largely ceases below 200 K (vertical line). We therefore conclude that the temperature evolution of the sample outside of this narrow temperature range can be neglected for the determination of the critical cooling rate. Note that the simulations do not show quantitative agreement with the crystalline fractions that we observe experimentally (Fig. 3c), which is likely due to our simplistic model for the temperature dependence of the nucleation rate. However, even if we vary the maximum nucleation rate in our model by several orders of magnitude, the narrow temperature interval in which crystallization occurs barely changes.

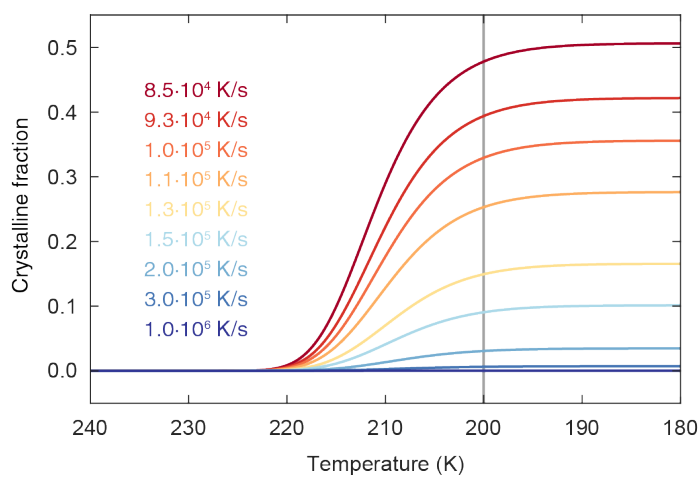

**Figure S1 | Simulation of the crystallization kinetics of the sample for different cooling rates.** The crystalline fraction of the sample is shown as a function of the sample temperature (details of the simulations in Ref. 4). Crystallization only sets in below 225 K and largely ceases below 200 K (grey vertical line).

**References**

(1) Olshin, P. K.; Bongiovanni, G.; Drabbels, M.; Lorenz, U. J. Atomic-Resolution Imaging of Fast Nanoscale Dynamics with Bright Microsecond Electron Pulses. Nano Lett. 2021, 21 (1), 612–618.

(2) Bongiovanni, G.; Olshin, P. K.; Drabbels, M.; Lorenz, U. J. Intense Microsecond Electron Pulses from a Schottky Emitter. Appl. Phys. Lett. 2020, 116 (23), 234103.

(3) Krüger, C. R.; Mowry, N. J.; Bongiovanni, G.; Drabbels, M.; Lorenz, U. J. Electron Diffraction of Deeply Supercooled Water in No Man's Land. Nat. Commun. 2023, 14 (1), 2812.

(4) Mowry, N. J.; Krüger, C. R.; Bongiovanni, G.; Drabbels, M.; Lorenz, U. J. Flash Melting Amorphous Ice. J. Chem. Phys. 2024, 160 (18), 184502.

(5) Krüger, C. R.; Mowry, N. J.; Drabbels, M.; Lorenz, U. J. Shaped Laser Pulses for Microsecond Time-Resolved Cryo-EM: Outrunning Crystallization during Flash Melting. J. Phys. Chem. Lett. 2024, 4244–4248.

(6) Hestand, N. J.; Skinner, J. L. Perspective: Crossing the Widom Line in No Man's Land: Experiments, Simulations, and the Location of the Liquid-Liquid Critical Point in Supercooled Water. J. Chem. Phys. 2018, 149 (14), 140901.

(7) Xu, Y.; Petrik, N. G.; Smith, R. S.; Kay, B. D.; Kimmel, G. A. Growth Rate of Crystalline Ice and the Diffusivity of Supercooled Water from 126 to 262 K; Proc Natl Acad Sci USA, 2016; Vol. 113.

(8) Stöckel, P.; Weidinger, I. M.; Baumgärtel, H.; Leisner, T. Rates of Homogeneous Ice Nucleation in Levitated H2O and D2O Droplets. J. Phys. Chem. A 2005, 109 (11), 2540–2546. https://doi.org/10.1021/jp047665y.

(9) Safarik, D. J.; Mullins, C. B. The Nucleation Rate of Crystalline Ice in Amorphous Solid Water. J. Chem. Phys. 2004, 121 (12), 6003–6010.

(10) Laksmono, H.; McQueen, T. A.; Sellberg, J. A.; Loh, N. D.; Huang, C.; Schlesinger, D.; Sierra, R. G.; Hampton, C. Y.; Nordlund, D.; Beye, M.; Martin, A. V.; Barty, A.; Seibert, M. M.; Messerschmidt, M.; Williams, G. J.; Boutet, S.; Amann-Winkel, K.; Loerting, T.; Pettersson, L. G. M.; Bogan, M. J.; Nilsson, A. Anomalous Behavior of the Homogeneous Ice Nucleation Rate in "No-Man's Land." J. Phys. Chem. Lett. 2015, 6 (14), 2826–2832.

(11) Murray, B. J.; Broadley, S. L.; Wilson, T. W.; Bull, S. J.; Wills, R. H.; Christenson, H. K.; Murray, E. J. Kinetics of the Homogeneous Freezing of Water. Phys. Chem. Chem. Phys. 2010, 12 (35), 10380.

(12) Riechers, B.; Wittbracht, F.; Hütten, A.; Koop, T. The Homogeneous Ice Nucleation Rate of Water Droplets Produced in a Microfluidic Device and the Role of Temperature Uncertainty. Phys. Chem. Chem. Phys. 2013, 15 (16), 5873.

(13) Stan, C. A.; Schneider, G. F.; Shevkoplyas, S. S.; Hashimoto, M.; Ibanescu, M.; Wiley, B. J.; Whitesides, G. M. A Microfluidic Apparatus for the Study of Ice Nucleation in Supercooled Water Drops. Lab. Chip 2009, 9 (16), 2293.